\documentclass{edm_article}

\usepackage{enumitem}
\usepackage{multirow}
\begin{document}
\title{%
% Score-Attended Feature Extraction for Deep Code Knowledge Tracing
% Code-DKT: Exploring When and How Domain-specific Features Can Improve Knowledge Tracing for Programming Tasks
% Code-DKT: A Model for Improving Knowledge Tracing for Programming Tasks with Domain-specific Features
Code-DKT: A Code-based Knowledge Tracing Model for Programming Tasks
}

% Submissions for EDM are double-blind: please do not include any author names or affiliations in the submission. 
% Anonymous authors:
% \numberofauthors{1}
% \author{
% Anonymous\\
%       \affaddr{Anonymous Institution}\\
%       \email{anonymous@anonymous.edu}
% }

\numberofauthors{1} %  in this sample file, there are a *total*
% of EIGHT authors. SIX appear on the 'first-page' (for formatting
% reasons) and the remaining two appear in the \additionalauthors section.
%
\author{
% You can go ahead and credit any number of authors here,
% e.g. one 'row of three' or two rows (consisting of one row of three
% and a second row of one, two or three).
%
% The command \alignauthor (no curly braces needed) should
% precede each author name, affiliation/snail-mail address and
% e-mail address. Additionally, tag each line of
% affiliation/address with \affaddr, and tag the
% e-mail address with \email.
%
%1st. author
\alignauthor Yang Shi, Min Chi, Tiffany Barnes, Thomas W. Price \\
      \affaddr{North Carolina State University}\\
      \affaddr{Raleigh, NC, USA}\\
      \email{\{yshi26, mchi, tmbarnes, twprice\}@ncsu.edu}
}

\maketitle

\begin{abstract}
Knowledge tracing (KT) models are a popular approach for predicting students' future performance at practice problems using their prior attempts. 
%Knowledge tracing (KT) models have seen rapid development in recent years by using prior student attempts at practice problems to accurately predict future performance. 
%In CS education, tracking student code submissions is more challenging, as they are more complicated, and different submissions can be all correct. 
%Though many innovations have been made in KT, most models including the state-of-the-art Deep KT (DKT) mainly leverage each student's response either as correct or incorrect, ignoring \emph{its content}.
%Despite many innovations, most KT models continue to treat each student's attempt as either ``correct'' or ``incorrect'', while ignoring the content of the student's response. 
%TB - replaced this sentence with the following one. In many domains such as programming education, however,   how students construct a solution attempt can provide valuable insights into inferring the skills they have mastered, which can be used to improve a KT model.
Though many innovations have been made in KT, most models including the state-of-the-art Deep KT (DKT) mainly leverage each student's response either as correct or incorrect, ignoring \emph{its content}.
%how the student constructs a solution attempt can be very informative for inferring the skills they have mastered, which may improve the performance of a KT model.
%In this work, we propose \textbf{Code-based Deep Knowledge Tracing (Code-DKT) model} that extends DKT by automatically extracting domain-specific features and selecting code features by using an attention mechanism.
In this work, we propose \textbf{Code-based Deep Knowledge Tracing (Code-DKT)}, a model that uses an attention mechanism to automatically extract and select domain-specific code features to extend DKT.
%we propose we a Code-based Deep Knowledge Tracing (Code-DKT) model, which augments DKT by  automatically extracting domain-specific features and by selecting code features using an attention mechanism. explore domain-specific feature extraction methods to improve KT model 
We compared the effectiveness of Code-DKT against Bayesian and Deep Knowledge Tracing (BKT and DKT) on a dataset from a class of 50 students attempting to solve 5 introductory programming assignments. Our results show that Code-DKT consistently outperforms DKT by $3.07-4.00\%$ AUC across the 5 assignments, a comparable improvement to other state-of-the-art domain-general KT models over DKT. Finally, we analyze problem-specific performance through a set of case studies for one assignment to demonstrate when and how code features improve Code-DKT's predictions.
%The decomposed results on different problems suggest that Code-DKT works better when multiple similar problems present in the dataset. We also use case studies to compare DKT and Code-DKT, further showing scenarios when code features help better. This paper implicates that code features would help knowledge tracing tasks in CS education using specifically designed mechanisms, and this conclusion could be potentially used in other similar domains. 
\end{abstract}

\keywords{Knowledge Tracing, Deep Knowledge Tracing, CS Education, Code Analysis, Deep Learning} % Replace with your own 3-5 keywords

\section{Introduction} % 1 page with abstract

% What is knowledge tracing

Modeling student knowledge to predict performance on future problems, called Knowledge Tracing (KT), is a fundamental feature of intelligent tutoring systems \cite{VanLehn2006}. %Improving these \textit{knowledge tracing} (KT) algorithms has been a core pursuit of the educational data mining (EDM) community. 
KT models enable tutoring systems to support mastery learning \cite{corbett1994knowledge}, select appropriate next problems \cite{ai2019concept}, provide help \cite{swamy2018deep}, and provide analytics to instructors \cite{piech2015deep}, all of which can improve learning.
%
%Knowing the learning status of student learning can bring multiple benefits. If the student is projected to be failing in their next attempt, it would be helpful to provide interventions. Data-driven methods have been introduced to solve this problem, as when students pass through courses they leave a trace of assignment submissions and their scores available. Researchers have leveraged the score and assignment traces to track student mastery of knowledge, and various statistical models are developed for this task, for example, Bayesian knowledge tracing (BKT) \cite{corbett1994knowledge}. 
%
KT models have  increased in complexity from the early 4-parameter Bayesian Knowledge Tracing (BKT) %that modeled ``guessing'' and ``slipping'' 
to modern models that train deep neural networks with tens of thousands of parameters using the latest deep learning innovations (e.g. attention \cite{xu2015show} and transformers \cite{vaswani2017attention}).
%BKT used parameters such as guessing or slip to model the learning process of students, assigning semantically meaningful, but limited parameters. More recently, deep learning models have been widely applied throughout different areas, largely increasing the parameter size to catch more information in datasets. These models have been applied to solve knowledge tracing problems in multiple publicly available datasets since 2015, when Piech et al. proposed the basic deep knowledge tracing \cite{piech2015deep}. 
%
This has led to improvement in KT model performance, especially for larger datasets, e.g. from ASSISTments \cite{selent2016assistments, gervet2020deep}.

The \textit{simplest version} of the KT problem uses only the sequence of: 1) which problems the student has attempted, and 2) whether or not each attempt was correct. While this makes KT models widely applicable across domains, this also omits a potential wealth of information about \textit{how} the student attempted each problem. Increasingly, ITS being built to support complex problem solving tasks, like programming in Snap \cite{price2017isnap} and in games \cite{peddycord2014generating}, logic proofs \cite{mostafavi2017evolution}, science inquiry \cite{leelawong2008designing} and language learning \cite{swartz2012intelligent}. In these domains, correctness may not provide enough information about student knowledge, varying significantly in the reasons both for incorrectness and correctness. In programming, for example, one incorrect attempt may have a minor syntax error while  another includes a clear misconception. Similarly, two different \textit{correct} answers could reveal dramatically different levels of concept mastery depending on their conciseness and the concepts used. Most KT models would treat all correct and all incorrect attempts identically. A domain-specific KT model, e.g. those for science by Rowe et al.  \cite{ROWE2017617}, might greatly improve KT performance.
Little work has investigated whether domain-general KT models can predict student success in programming, or how domain-specific features might  improve performance.
In this paper, we explore when and how features extracted from students' submitted code can improve a KT model for programming. To do so, we introduce a novel code-based  deep knowledge tracing (Code-DKT) model, which uses the code2vec model \cite{alon2018general} to learn a meaningful representation of student code, and combines this with Deep Knowledge Tracing (DKT)  \cite{piech2015deep} to track student progress. Specifically, student code submissions are represented with abstract syntax trees, and split into multiple code paths \cite{alon2018general} (explained in Section~\ref{sec:method}). We assign the importance of different code paths by learning weights guided by the scores students received for the current and past submissions. We compared the performance of Code-DKT with baseline BKT and DKT models on a dataset of 50 introductory programming problems from 410 students, across 5 assignments. Our experiments show that the Code-DKT model is able to consistently improve DKT's performance by 3.07-4.00 percentage points in AUC. This improvement is comparable to that of other modern KT models over DKT (2-4\%) \cite{pandey2019self,shin2021saint+}, suggesting that domain-specific features may be just as important as model structure. Finally, we investigate one assignment through 3 case studies to explore the mechanisms by which code features may improve the model, and when they are most useful. We also show that Code-DKT outperforms more naive code-feature models. Overall, this paper makes three contributions: 1) the Code-DKT model, which extends DKT for programming tasks; 2) evidence that Code-DKT outperforms both domain-general models and naive code-feature models; and 3) evidence of when and how Code-DKT's code features improve model performance.

%To sum up, this paper answers three research questions: \textbf{RQ1} How well do existing KT methods work in the domain of programming? \textbf{RQ2} Can existing KT models be easily extended to incorporate code information? How does Code-DKT -- a programming-specific KT model that incorporates code structure -- compare to DKT? \textbf{RQ3} When and how do code features improve model performance?

% Potential contributions:
% - Showing that, despite what you may have thought, domain specific features do not necessarily improve model performance by much (can we show some of the "why" here?
% - A novel approach for KT in programming domains that _does_ improve model performance by a large margin
% - Show through a breakdown of first/non-first attempts, as well as case studies, how the code features improve the model: primarily by improving it's ability to predict within-problem based on how close students are to the correct solution.

\section{Related Work} % 1.75 pages
In this section, we present related work on knowledge tracing, student modeling in computer science education, and deep learning models for code/programs.

\subsection{Knowledge Tracing}
Knowledge tracing (KT) models student knowledge as they solve problems  to predict future performance. In KT, problems are labeled with needed skills (i.e. knowledge components, KC)   \cite{vanlehn1988student}, the skill or q-matrix can be learned from data \cite{barnes2005q}, or the problem ID can be used instead. In Bayesian Knowledge Tracing (BKT), the most popular KT method \cite{corbett1994knowledge}, a simple Bayesian model is built to model student knowledge using parameters for guess (getting a problem right when a skill is not known), slip (getting it wrong when known), and transition from unlearned to learned after practicing. These parameters are learned from prior students' problem sequences, and then used to predict future performance. Researchers have improved BKT performance, for example, by calculating the bound or prior distribution of parameters \cite{d2008more}, adding a priori estimates of student learning \cite{pardos2010modeling}, or integrating speed factors \cite{yudelson2013individualized}. 

A number of innovations have improved \textit{domain general} KT, without using additional features from student's work (only the correctness of each problem attempt). With the development of machine learning technologies and increasingly available large datasets, models based on deep learning have been proven more effective, especially with enough available data \cite{gervet2020deep}. Piech et al. introduced deep knowledge tracing (DKT), using recurrent neural networks (RNN) to predict a student's knowledge of each skill (or problem) after each problem attempt, and to learn the relationships among skills automatically \cite{piech2015deep}. As our work is based on this model, we will discuss the details of the model in Section~\ref{sec:method}. %The success of DKT attracted more related efforts on deep learning for knowledge tracing. 
Some recent advances in deep learning for knowledge tracing focus on model structure, including SAKT and SAINT. Self-attentive knowledge tracing (SAKT) \cite{pandey2019self} added a self-attention mechanism \cite{xu2015show} to DKT, while Separated Self-Attentive Neural Knowledge Tracing (SAINT) \cite{choi2020towards} later integrated a transformer (a type of deep neural network which has been successfully applied in text and image processing areas) into a knowledge tracing model \cite{vaswani2017attention}. Both of these models have outperformed DKT, especially on large datasets such as EdNet \cite{choi2020ednet}, e.g. by 2\% AUC.

While these innovations have improved KT performance, often using complex networks and larger datasets, the datasets used generally only indicate whether a student's attempt was \textit{correct}, but not the \textit{content} of a student's answer or their \textit{process} for achieving it, and the models therefore do not use this information. However, researchers have incorporated \textit{other types} of  information into deep models, such as course prerequisites or the relationships among problems. For instance, Chen et al. attached prerequisite information in the DKT modeling process for a more accurate prediction\cite{chen2018prerequisite}. The prerequisite concepts were modeled as graph matrices (as done by Wang et al. \cite{wang2016using}), serving as an additional input to knowledge tracing models, similar to skill or q-matrices that can also be learned from student data \cite{barnes2005q}. On the other hand, Ghosh et al. introduced attentive knowledge tracing (AKT) \cite{ghosh2020context}. They introduced a decay parameter to explicitly reduce the impact of distant problems, and at the same time used a Rasch model \cite{rasch1993probabilistic} to incorporate problem contexts, then embedding the differences among the problems. Student information can also be used for knowledge tracing models. Educational priors such as a learning or forgetting curves can be integrated into deep knowledge tracing models \cite{chen2017tracking}. The closest such models come to incorporating students' solution processes is including information about how \textit{fast} students solved a problem. Yudelson et al. \cite{yudelson2013individualized} added speed factors into BKT, and similar temporal information can also improve the performance of deep models (e.g. \cite{shin2021saint+}).

None of the above work have used the student response information, besides submission correctness, in their models. This could be partially because of the simplicity of the problems. Most of them are true or false, multiple choice problems, or short answer problems. The availability of the exercise data is also limited, as some datasets only contains a sequence of binary correctness scores from students. Recent work (e.g. EKT, EERNN \cite{su2018exercise, liu2019ekt}) used a joint embedding of \textit{exercise} text and response correctness, combining the exercise text embedding together with student scores to represent student individualized submissions. This achieved better performance than the other models without using this information. However, these models only use \textit{problem information}, but no information about the \textit{students' answer} beyond binary correctness information. This suggests an opportunity to create improved, \textit{domain-specific} KT models in areas such as programming, math, science or writing, where students' answers include complex written responses or structured problem-solving steps. Recent work has incorporated such problem-specific data in deep learning approaches used to adapt pedagogical policies for tutoring in logic \cite{maniktala2020extending}, probability \cite{zhou2019hierarchical}, or predict performance in programming \cite{mao2019one} but generally have not been used to built KT models in these domains. In the domain of programming education, for example, students' code submissions contain rich information on the state of their current knowledge. As proposed in this paper, more structural information could be extracted from student code submission to infer students' learning status of certain concepts. We use these code features to make better knowledge tracing models.
% DKT serves as as the base model due to its simplicity, as integrating code feature efficiently into other more advanced deep learning models may require extra design, and we mainly focus on a proof-of-concept that the use of code features can improve the performance of KT models.

\subsection{Student Modeling in CS Education}

% Other KT or knowledge component modeling work in CS education, that may have used programs
%In recent years, student modeling studies in CS Education (CSEd) have started to use student source code as input into the modeling process. 
Researchers in CS education have explored ways to model student source code for intelligent tutoring. In 2011, Jin et al. proposed that a linkage representation that reflected code structure could be used for programming hint generation \cite{jin2011towards}. In 2014, Yudelson et al. extracted code features from a MOOC on introductory Java programming to explore code recommendation methods \cite{yudelson2014investigating}. Their work focused on using a combination of problem correctness and extracted code features to predict student success, and use this prediction to recommend an appropriate next problem to a student. While they did not evaluate their model on a KT task per se, their approach of extracting atomic code features is somewhat similar to our TFIDF baseline (Section~\ref{sec:baselines}). 
Another work from Rivers et al. used code features for student learning curve analysis and attempted to directly extract meaningful knowledge components (and whether they were successfully applied) from student code \cite{rivers2016learning}. In their work, student code submissions are represented as abstract syntax trees (ASTs), with the node types of ASTs (e.g. \texttt{for}, \texttt{if}) representing knowledge components (KCs). The error rate curves (referred to as ``learning curves") were plotted over time, visualizing the mastery of different KCs. They showed that while code-based KCs produced well-fitting curves, others did not. While this suggests the possible validity of AST-based KC extraction, the work did not directly evaluate the utility of these KCs for knowledge tracing.
Like our current work, Wang et al. showed that incorporating structural code features can improve DKT  for a single problem from a large ``hour of code" (HoC) dataset \cite{wang2017learning}. However, this HoC exercise has a very simple solution, so their results may not generalize. Additionally, their features were learned in an unsupervised way from ASTs, while our approach learns an embedding from the data.  

Code features have also been used in tasks other than KT as well, such as common bug identification in student code. Traditionally, experts manually examined student code to identify common bugs in different student levels and programming languages, such as Java \cite{truong2004static} or block based programs \cite{hermans2016code}. However, manual examination is expensive for large-scale and quantitative studies. More advanced work takes advantage of the growing size of datasets, and used data-driven methods to find bugs in student code submissions. For example, Choi et al. used simple machine learning methods to detect malicious code in code by using simple feature extraction methods such as counting neighboring tokens in code text (n-gram). With the recent advance of computational power and even bigger datasets, more deep learning methods have emerged. These methods focused on developing deep neural network methods to extract structural information for automatic student bug detection. For example, Gupta et al. used a matrix to represent the ASTs of student code to localize student code submissions \cite{gupta2019neural} in a large dataset (270K samples). For smaller sized dataset, Shi et al. evaluated the bug detection performance with the help of semi-supervised learning \cite{shi2021more}, and have also shown that unsupervised learning is possible with the help of experts \cite{shi2021toward}. All these methods reported better performance than traditional data-driven models on their tasks, showing the feasibility of similar usage on KT tasks.

% While it hasn't been used much for KT specifically, code features are often used in predictive tasks in programming education, so it's reasonable to believe these would improve a kt model.

While we focus on using student code submissions to extract features for student programming KT tasks, other less complicated approaches exist. Original programming tutors such as ACT \cite{corbett1997student} and Lisp tutor \cite{anderson1985lisp} decompose computational problems into small steps and let students make choices. This facilitates the KT tasks, as in these datasets, student submissions are simple multiple choices. However, with the development of newer Intelligent Tutoring Systems (ITSs), more systems provide intelligent support to students' written code. This provides better practice for students, but also makes knowledge tracing in computer science a more challenging task. Our paper aims at extracting code features for KT tasks in these new datasets.

\subsection{Deep Code Learning}

% How traditionally machine learning algorithms work for code tasks
Besides code feature extraction in the CS education domain, programming code has also been analyzed with data-driven models in software engineering research. For example, Allamanis et al. used neighboring tokens in source code (n-grams) to represent programming code, borrowing methods from natural language processing studies to predict method names in big code datasets \cite{allamanis2013mining}. Later work further explored extracting features from code structure, such as Raychev et al. who used decision trees to model programming code, making probabilistic predictions on the types of nodes in AST \cite{raychev2016probabilistic}. However, these simple structural approaches are often outperformed by  newly developed deep learning models, especially when applied to big datasets.

% How deep neural networks understand code, code representations

Deep neural networks have been applied in the software engineering domain, and achieved better performance than traditional data-driven methods. For example, Allamanis et al. used convolutional neural networks (CNNs) to classify code functions \cite{allamanis2016convolutional}; Mou et al. reworked the CNNs to an AST version, using the parent-children direction information in tree representations. Both methods greatly improved method classification tasks on classical machine learning models. Another recent model, code2vec, outperformed these models. Alon et al. designed this model, which leverages nodes and traversal paths in the ASTs to represent programs \cite{alon2019code2vec}. In their work, the leaf nodes of the ASTs are selected to represent the semantic information about the code. In addition, as there is a path through the AST from every leaf node to any other leaf node, this path is extracted to represent the code's structural information. The traversal paths together with the corresponding leaf nodes serve as the basic units of a representation of code \cite{alon2018general}. The code2vec model calculates the weight of each code path using an attention mechanism \cite{xu2015show} to automatically classify function names. Code-DKT's code extraction component is based on the code2vec model, but adds score to the attention mechanism to assign weights to code paths \cite{alon2018general} for predictions.

We chose code2vec to represent student code in DKT due to its recent successes for modeling code, and its attention mechanism. The attention mechanism learns weights for different features, allowing the model to directly use score information to select the most predictive code paths. Future work could investigate other code representations such as ASTNN, which has also been applied to make predictions from student code \cite{mao2021knowing}, or more recent advances such as CodeBERT \cite{feng2020codebert}.

\section{Method} % 1.5 pages
\label{sec:method}

\begin{figure}
\begin{center}
\includegraphics[width=0.30\textwidth]{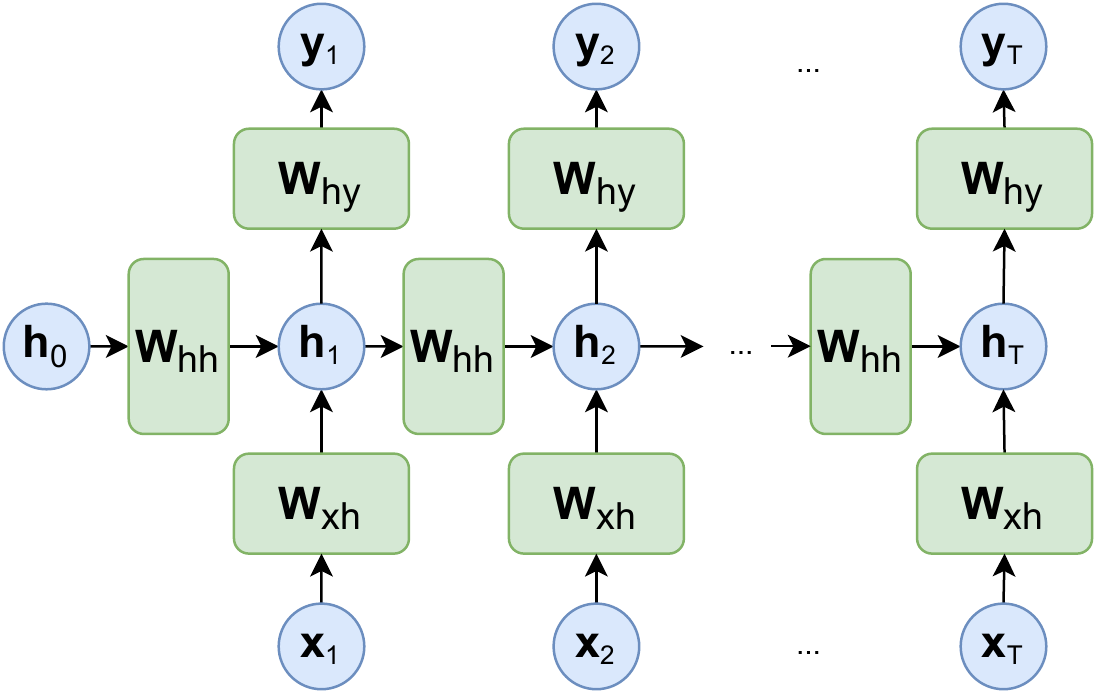}
\caption{\label{fig:RNN} Recurrent neural network structure.}

\end{center}
\end{figure}

\begin{figure*}
\centering

\begin{minipage}[b]{0.3\textwidth}
\includegraphics[width=\textwidth]{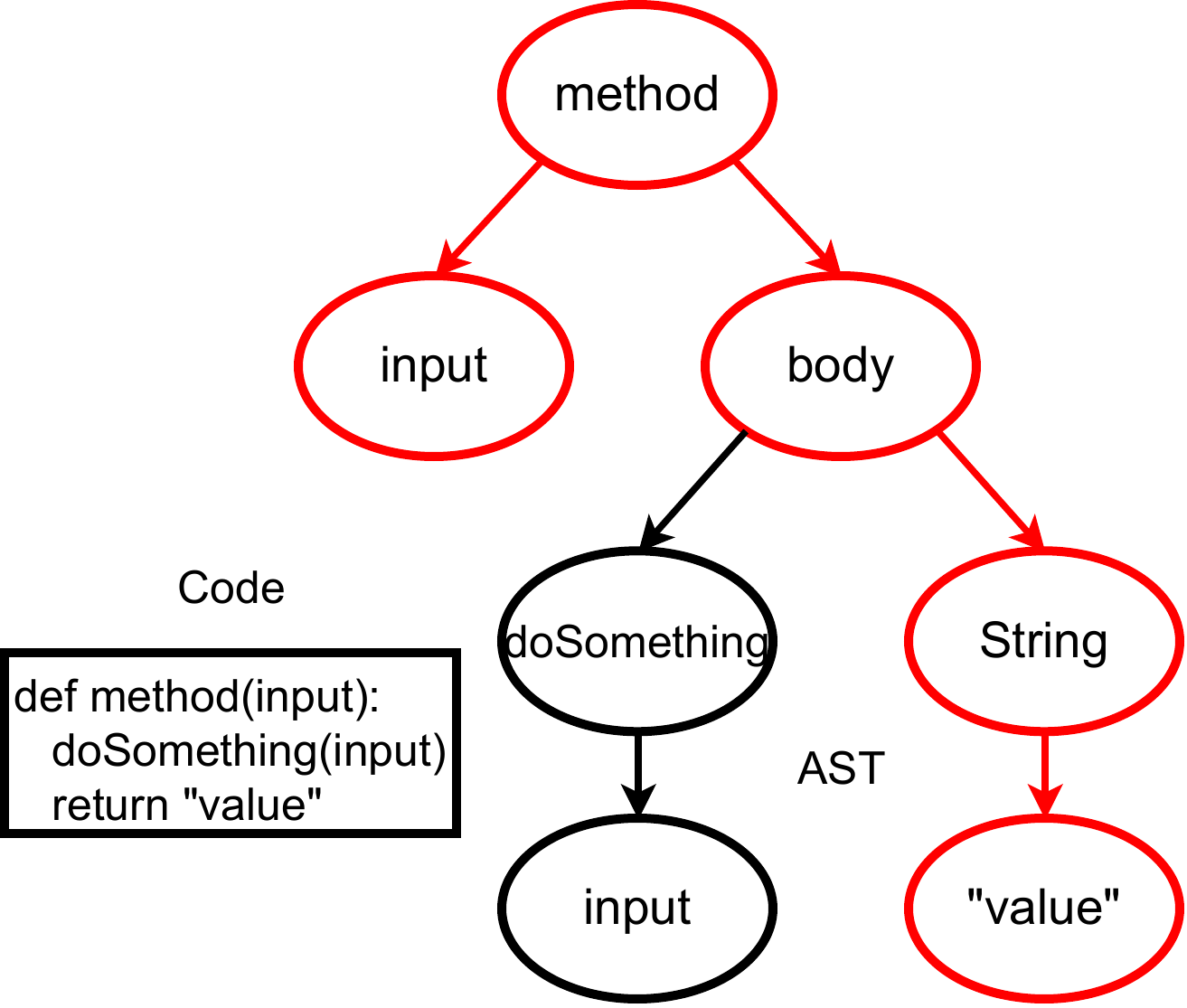}
\caption{\label{fig:AST} A simple AST where red nodes and edges represent a leaf-to-leaf path from \texttt{input} to \texttt{"value"}.}

\end{minipage}
\hfill
\begin{minipage}[b]{0.64\textwidth}
\includegraphics[width=\textwidth]{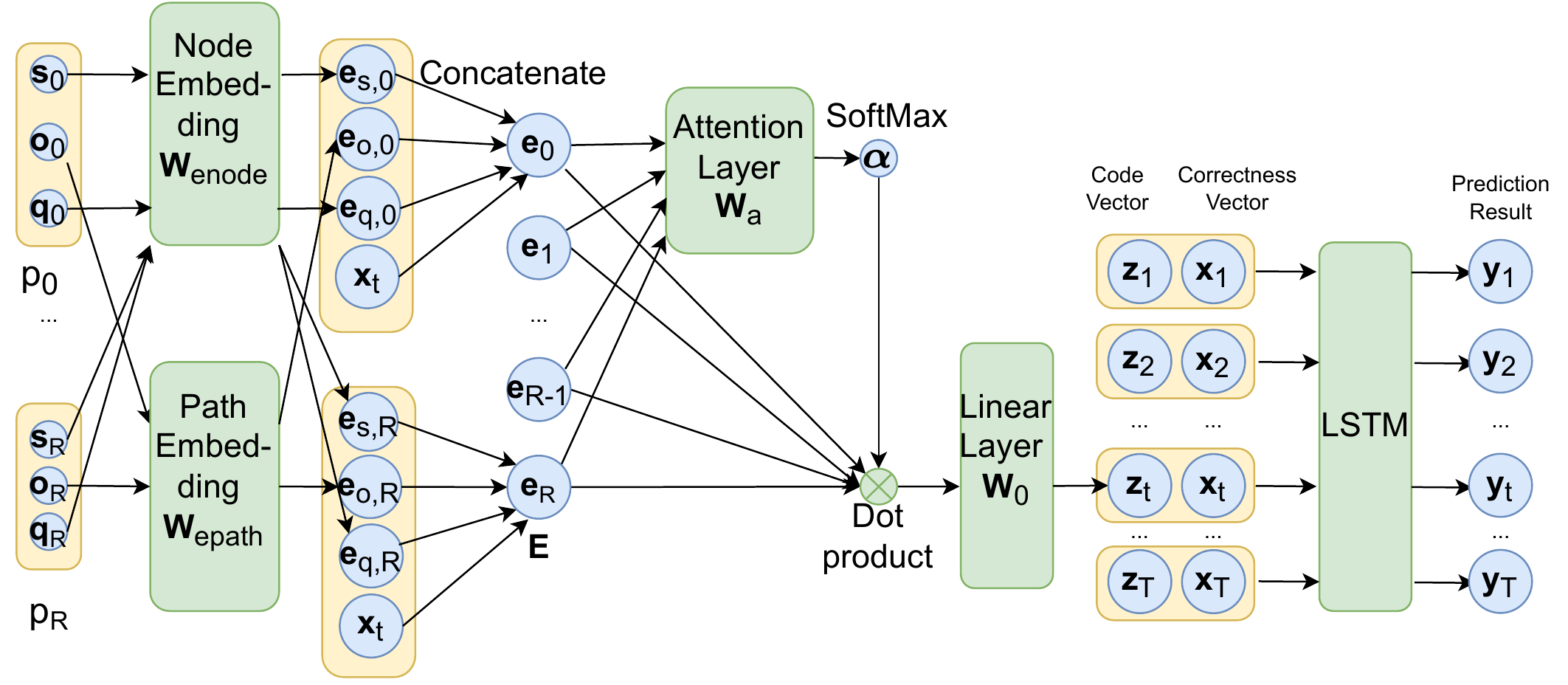}
\caption{\label{fig:Code-DKT}Code-DKT model structure.}

\end{minipage}
% \vspace{-0.2cm}
\end{figure*}

\textbf{Problem Definition}: Knowledge tracing (KT) tasks model a prediction problem: Given the history of a student's attempts at various KCs/problems, the model predicts if the student will succeed on their next attempt \footnote{We use problemIDs for KCs in this work}. Specifically, we define each student attempt $\mathbf{x}_t$ at time $t$ as $(q_t, a_t, c_t)$, where $q_t$ is the problem ID, $a_t$ is the correctness, and $c_t$ is the program code submitted for this attempt. Historically, KT algorithms have only utilized $q_t$ and $a_t$, and in this work we extend the input sequence to include $c_t$. At each timestep $T$, the model is given the $T$-length student attempt sequence $\mathcal{S}_T = \{(q_1, a_1, c_1),$ $ (q_2, a_2, c_2),...,(q_T, a_T, c_T)\}$, and it predicts whether the student's next attempt ($T+1$) on a given problem ($q_{T+1}$) will be correct ($a_{T+1}$). Note that students may attempt problems multiple times, and the model will make a prediction at each attempt.

Our proposed Deep Code Knowledge Tracing (Code-DKT) model integrates deep knowledge tracing (DKT) \cite{piech2015deep}) with the code2vec classification algorithm \cite{alon2019code2vec}. In this section we introduce the DKT model and how we enhance it with code feature extraction and selection.

\subsection{Deep Knowledge Tracing}
\label{sec:dkt}
Deep knowledge tracing uses a recurrent neural network (RNN) structure to learn the probability that a student will make a correct attempt on a subsequent problem. In the original implementation of DKT, the authors also implemented a version of DKT using a long short term memory (LSTM) model \cite{hochreiter1997long}, which is widely perceived as an advancement over RNNs. For simplicity, we explain DKT using an RNN model; we performed DKT using both RNNs and LSTMs. In the experiments, the LSTM version yielded higher performance\footnote{See the appendix of \cite{piech2015deep} for the LSTM DKT equations.} (see performance comparison in Section~\ref{sec:ablation}). We chose DKT as our baseline model, to compare with and to extend, as it is a commonly used baseline in other more recent KT papers \cite{pandey2019self, shin2021saint+}. Further, its LSTM structure makes it straightforward to extend with code features and to directly evaluate those features' contributions. Some recent models have outperformed DKT, but only by about 2-4\% AUC \cite{pandey2019self, shin2021saint+}, suggesting that DKT is still representative of modern deep KT models.

\textbf{Model Input}: %The RNN model takes a sequence of vectors as input, and uses this to make predictions. In DKT, this sequence is $\mathcal{S}_T$, the sequence of attempts (including which problem was attempted and it's correctness), which are one-hot encoded into binary vectors. Specifically, given a $T$-length sequence of $\mathcal{S}$, where we have $M$ problems in the dataset, each problem-correctness pair $\{q_t, a_t\}$ is one-hot encoded into a binary vector $\mathbf{x}_t$ of size $2M$, where $x_{q_t + M(1-a_t)}$ is set to 1, and the other bits are set to 0. For example, if we have $3$ problems in the dataset, if the student succeeds on the first problem, $\mathbf{x}=\{1,0,0,0,0,0\}$, and if they fail on the first problem, $\mathbf{x}$ is assigned as $\{0,0,0,1,0,0\}$ to represent their failure.
For each student, DKT (RNN) takes as input a sequence $\mathcal{S} = \{\mathbf{x}_1, \mathbf{x}_2, ... \mathbf{x}_T\}$ of $T$ attempt vectors $\mathbf{x}_t$.  With $M$ problems, each attempt consisting of problem-correctness pair $\{q_t, a_t\}$ at time $t$, is one-hot encoded into a binary vector $\mathbf{x}_t$ of size $2M$, where $x_{q_t + M(1-a_t)}$ is set to 1, and the other bits are set to 0. For example, with $M=3$, for student success on problem 1, $q_t=1, a_t=1$, so $x_{1+3(1-1)}=1$, so $\mathbf{x}=\{1,0,0,0,0,0\}$, and failure on problem 1 $q_t=1, a_t=0$, so $x_{1+3(1-0)}=1$, so $x_4$ is set to one, and $\mathbf{x}$ is  $\{0,0,0,1,0,0\}$.
%
% The correctness embedding vector $\mathbf{x}_t$ only has $1$ element as $1$, while other elements are all $0$s. If the submission is correct, where $a_t = 1$ on the problem $q_t$, the element indexed $q_t$ is assigned as $1$ on $\mathbf{x}_t$; if the submission is incorrect ($a_t = 0$ on the problem $q_t$), then the element indexed $q_t+Q$ is assigned as $1$ on $\mathbf{x}_t$. 
%
%Every student thus has a DKT input sequence $\mathcal{S} = \{\mathbf{x}_1, \mathbf{x}_2, ... \mathbf{x}_T\}$.

\textbf{Model Structure}: The RNN version of DKT  maps each input sequence $\mathcal{S}_T$ into an output sequence of predictions $\mathcal{Y} = \{\mathbf{y}_1, \mathbf{y}_2, ..., \mathbf{y}_T\}$ with a set of hidden states $\mathbf{h}_1, \mathbf{h}_2, ... \mathbf{h}_T$. More specifically, as illustrated in Figure~\ref{fig:RNN}, this process is defined as:
$$\mathbf{h}_t = \mathrm{tanh}(\mathbf{W}_{xh}\mathbf{x}_t+\mathbf{W}_{hh}\mathbf{h}_{t-1}),$$
$$\mathbf{y}_t = \sigma(\mathbf{W}_{hy}\mathbf{h}_t).$$

In the equations, element-wise operators $\mathrm{tanh}(\cdot)$ and $\sigma(\cdot)$ are activation functions of the network, introducing non-linearity to the network. The parameters learned in the network are $\mathbf{W}_{xh}$ which transforms input $\mathbf{x}_t$ into the hidden space, $\mathbf{W}_{hh}$ which fuses the hidden state $\mathbf{h}_{t-1}$ from the prior input with the current hidden state $\mathbf{h}_{t}$, and $\mathbf{W}_{hy}$ which translates the hidden state $h_t$ into an output. In both equations, the bias terms are omitted for simplicity, and the $\mathbf{h}_0$ is the initial hidden state, the zero-vector.

\textbf{Model Output}: The output sequence $\mathcal{Y}$ contains prediction vectors $\mathbf{y}_t$, sized $M$. Every element of the vector represents the probability of the student making a correct submission on corresponding problems in their next attempt. Note that while the model makes predictions for each problem at each timestep $t$, only the value for the next attempted problem $q_{t+1}$ is used during training and evaluation.

% \begin{figure}
% \begin{center}
% \includegraphics[width=0.33\textwidth]{Figures/AST.pdf}
% \caption{\label{fig:AST} An AST representing some simple code, where red nodes and edges in the tree represents a code path example starting from \texttt{input} to \texttt{"value"}.}
% \end{center}
% \end{figure}
\subsection{Deep Code Knowledge Tracing}
\label{sec:code-dkt}

% \begin{figure*}
% \begin{center}
% \includegraphics[width=0.65\textwidth]{Figures/DCKT.pdf}
% \caption{\label{fig:Code-DKT}Code-DKT model structure.}
% \end{center}
% \end{figure*}

We extend DKT into Deep \textit{Code} Knowledge Tracing (Code-DKT), by using the code2vec \cite{alon2019code2vec} representation of student code attempts, $c_t$, along with problem and correctness information. %Our approach is based on the code2vec \cite{alon2019code2vec} code classification model.

\textbf{Code Representation}: Abstract syntax trees (ASTs) are used to represent the hierarchical structure of code, for example with a node for a function (\texttt{method}) with children representing the function's parameter (\texttt{input}) and body (\texttt{body}). AST leaf nodes often correspond to literal values or identifiers. Code-DKT extends the code2vec model for code classification, which encodes an AST using a set of \textit{leaf-to-leaf paths} throughout the AST. For example, in Figure~\ref{fig:AST}, a path from the leaf node  \texttt{input} to the leaf node \texttt{"value"} (highlighted red in the example) consists of the nodes: [\texttt{input}, \texttt{method}, \texttt{body}, \texttt{String}, \texttt{"value"}]. Given an AST, code2vec extracts a set of leaf-to-leaf paths, as explained below.

\textbf{Model Input}: Since a deep learning model cannot operate directly on code paths, the Code-DKT must next convert this code-path representation of the AST into a binary vector.  A student's code submission $c_t$ at time $t$ is represented as $\{p_0, p_1, ... ,p_R\}$ where there are in total $R$ randomly selected code paths in $c_t$. Every $p_r$ has three components, namely the starting node of the code path $\mathbf{s}_r$, the textual representation of the full path $\mathbf{o}_r$, and the ending node $\mathbf{q}_r$, which are each one-hot encoded as binary vectors. For instance, for the example in Figure~\ref{fig:AST}, $\mathbf{s}_r$ is \texttt{input}, $\mathbf{o}_r$ is a text string: \texttt{input|method|body|String|value}, and $\mathbf{q}_r$ is \texttt{value}.

\textbf{Model Structure}: Rather than using a \textit{static} vector representation of students' code, Code-DKT \textit{learns} an optimal embedding of student code.
%using the training data. 
The detailed Code-DKT model structure is shown in Figure~\ref{fig:Code-DKT}. This initial structure is drawn from code2vec. The nodes for each of $R$ code paths in $c_t$ ($c_t$ has in total $R$ paths), including starting and ending nodes $(\mathbf{s}_r, \mathbf{q}_r)$ and paths $\mathbf{o}_r$ for a single path $r$, are respectively embedded by the node embedding matrix $\mathbf{W}_{enode}$ and the path embedding matrix $\mathbf{W}_{epath}$. Both matrices are randomly initialized with a Gaussian distribution, but they are later updated during model training. The Code-DKT model structure then diverges somewhat from code2vec, to account for the specific needs of the KT problem. Specifically, the three embedded vectors representing $c_t$ are concatenated with the problem-correctness vector $x_t$ from DKT (introduced in Section~\ref{sec:dkt}). This serves as a numerical representation of $(q_t, a_t, c_t)$. For a single code path $p_r$, this process is accomplished with embeddings for the start node ($\mathbf{e}_{s,r}$), path ($\mathbf{e}_{o,r}$), and end node ($\mathbf{e}_{q,r}$):
$$\mathbf{e}_{s,r} = \mathbf{W}_{enode}\mathbf{s}_r; \mathbf{e}_{o,r} = \mathbf{W}_{epath}\mathbf{o}_r; \mathbf{e}_{q,r} = \mathbf{W}_{enode}\mathbf{q}_r,$$
$$\mathbf{e}_{r} = [\mathbf{e}_{s,r}; \mathbf{e}_{o,r}; \mathbf{e}_{q,r}; \mathbf{x}_{t}].$$

\textbf{Score-Attended Path Selection}: Code-DKT now has an numerical representation of a single attempt: a set of $R$ embedded vectors, $\mathbf{e}_{r}$, one for each code path in $c_t$. Note that the embedding,  $\mathbf{e}_{r}$ not only includes the code information, but also the current correctness score information $\mathbf{x}_t$ at the submission $t$. However, not all parts of a student's code are relevant, and thus not all code paths $\mathbf{e}_{r}$ are important for predicting a student's future success. Therefore, the model uses an attention mechanism to identify how much weight to give to each of these paths. The embedding vectors $\mathbf{E} = \{\mathbf{e}_{0}, \mathbf{e}_{1},..., \mathbf{e}_{R}\}$ are multiplied by the attention matrix $\mathbf{W}_a$ to get $R$ scalars $a_{0}, a_{1},..., a_{R}$, representing the importance (commonly known as the ``attention'') of each of the code paths. The importance $a_r$ uses a SoftMax mechanism for normalization, having $1$ as the sum.
%\footnote{The SoftMax operator is defined as $\mathrm{SoftMax}(\mathbf{a}) = \frac{e^{a_i}}{\sum_{i=1}^Re^{a_i}}$ for a vector $\mathbf{a}$ sized $R$.} 
This process is formulated as:
$$\mathbf{\alpha} = \mathrm{SoftMax}(\mathbf{E}\mathbf{W}_a)$$
$$\mathrm{SoftMax}(\mathbf{a}) = \frac{e^{a_i}}{\sum_{i=1}^Re^{a_i}}$$
where each elements $\alpha_r$ in $\mathbf{\alpha} = \{\alpha_1, \alpha_2, ..., \alpha_R\}$ are the calculated weights for the code path $p_r$. Finally, Code-DKT weights each code path $\mathbf{e_i}$ by its attention $\alpha_i$, and sums them together, giving a weighted average: a single vector representing the important parts of the code. The weighted average vector is then multiplied by a matrix $\mathbf{W}_0$ to get the code vector $\textbf{z}$, representing features extracted from code submissions, as in equation: 
$$\mathbf{z} = \mathbf{W}_0(\sum_{i=1}^R\alpha_i\mathbf{e}_{i}).$$

In a sequence of $T$ student attempts, Code-DKT produces $T$ code vectors $\{\mathbf{z}_1, \mathbf{z}_2, ... ,\mathbf{z}_T\}$. The code vectors are concatenated with the correctness vectors $\{\mathbf{x}_1, \mathbf{x}_2, ... ,\mathbf{x}_T\}$ as the input to the final LSTM (as in DKT), giving the predictions $\{\mathbf{y}_1, \mathbf{y}_2,$ $ ... \mathbf{y}_T\}$. Even though $x_t$ was already used to produce $\mathbf{z}_t$, this final concatenation ensures the Code-DKT model has direct access to the student correctness score information. %, rather than inferring them through the code feature selection network.

\section{Experiments} % 1.5 page
We designed an experiment to evaluate 3 research questions about student modeling in the domain of programming:

\begin{itemize}[noitemsep,topsep=0pt,parsep=0pt,partopsep=0pt]
    \item[\textbf{RQ1}] How effective are domain general KT approaches (DKT, BKT) on our programming dataset?
    \item[\textbf{RQ2}] How can features derived from students' code be used to improve KT models?
    \item[\textbf{RQ3}] When are these code features most useful, and how can they lead to improved predictions?
\end{itemize}

% Dataset, distribution
\subsection{Dataset \& Experiments Setup}
Our study uses a dataset of an introductory Java programming class at a large, university in the US, collected in Spring 2019, stored in the ProgSnap2 format \cite{price2020progsnap2}. The dataset includes work from $410$ students on $50$ problems divided over $5$ assignments. These were completed throughout the semester as homework, with each assignment focusing on a specific topic (e.g. conditionals, loops). For these problems, typical solutions ranged $10$ to $20$ lines of code. Students tended to make multiple submissions before succeeding finally, and $23.68\%$ of the attempts were correct. Student code was automatically graded using test cases, and We treated a submission as correct (1) only when all test cases passed, and incorrect (0) otherwise.

For each assignment, students were then split into training and testing sets with a ratio of $4:1$.  One quarter of the training data were used for hyperparameter tuning and validation (see below). Then, we trained the model on the whole training dataset, and tested on the holdout test dataset, repeating this process 10 times to account for model variation (e.g. due to random initialization). All deep learning models were implemented using the PyTorch\cite{NEURIPS2019_9015} library, and our BKT implementation was pyBKT \cite{badrinath2021pybkt}.

\subsection{Hyperparameter Tuning \& Optimization}
For hyperparameter tuning, we split the training data into training and validation sets, and created a model with each possible set of hyperparameters (described below), and calculated AUC performance on the validation dataset. We repeated this process 100 times and chose the hyperparameter setting with the best average validation performance to use in testing/evaluation. Specifically, we selected the embedding size of code feature extraction as $300$, from a range of ($50$, $100$, $150$, $300$, and $350$); learning rate was selected as $0.0005$ from a range of ($0.00005$, $0.0005$, $0.005$, $0.01$); the training epochs were set at $40$ to save training time while keeping the best prediction results, selected from a range of ($20$, $40$, $100$). All other parameters were defaulted as the original settings of code2vec and DKT. We fixed the longest length of student attempts at $50$ to filter extra long submission traces from students. In cases where more than $50$ attempts were submitted, we used the last $50$ submissions, assuming the latest submissions were more useful.

As the models were deep neural networks, we used binary cross entropy as a loss function to track the difference between the ground truth and predicted probabilities. 
%Binary cross entropy is the default loss function used by DKT, and the loss for a single student submission is calculated as:
%$$l = y_{t+1}\mathrm{log}(\hat{y}_t)+(1-y_{t+1})\mathrm{log}(1-\hat{y}_t),$$ where $y_{t+1}$ denotes the ground truth of student next submission correctness, and $\hat{y}_t$ specifies the predicted probability on the problem of next attempt, given current code and submission trace as input. 
The models used back propagation to update weight matrices (parameters), using the Adam optimizer \cite{kingma2015adam}, which is also a default for code2vec and DKT.\footnote{Repository: https://github.com/YangAzure/Code-DKT}

%Parameters

\begin{table}[]
\centering
\caption{\label{table:assignment} Performance Comparison on all assignments.}
\begin{tabular}{|c|c|c|c|c|c|}
\hline
Model &  A1       & A2       & A3       & A4       & A5       \\ \hline
DKT          & 71.24\% & 73.09\% & 76.84\% & 69.16\% & 75.14\% \\ \hline
Code-DKT         & 74.31\% & 76.56\% & 80.40\% & 72.75\% & 79.14\% \\ \hline
\end{tabular}
\end{table}

\begin{table}[]
\centering
\caption{\label{table:overall_comparison} Overall and the first attempt performance of all models on assignment A1.}
\begin{tabular}{|c|c|c|}
\hline
 \begin{tabular}[c]{@{}c@{}}Models\\ AUC (STD)\end{tabular} & \begin{tabular}[c]{@{}c@{}}Overall\\ \end{tabular} & \begin{tabular}[c]{@{}c@{}}First Attempts\\ \end{tabular} \\ \hline
Code-DKT                         & \textbf{74.31\%} (0.90\%)                                      & \textbf{75.74\%} (0.69\%)                                                   \\ \hline
DKT-TFIDF                    & 69.94\% (0.88\%)                                              & 72.77\% (0.79\%)                                                   \\ \hline
DKT-Expert                   & 69.52\% (0.68\%)                                              & 69.53\% (0.72\%)                                                   \\ \hline
DKT                          & 71.24\% (2.54\%)                                              & 72.26\% (3.69\%)                                                   \\ \hline
BKT                          & 63.78\% (4.68\%)                                              & 50.22\% (2.86\%)                                                   \\ \hline
\end{tabular}
\end{table}

\subsection{Baselines}
\label{sec:baselines}
%models compared, use cases of codes and heatmaps

We compare the performance of Code-DKT to DKT, BKT, and two modified DKT methods: DKT-TFIDF adding data-driven features, and DKT-Expert adding expert features. Specifically,  DKT-TFIDF uses TFIDF,  a data-driven feature that counts the term frequency (TF) of tokens (variables, functions, and operations, etc.) in code text, and forms a frequency vector for every term. This frequency is multiplied by the inverse document frequency (IDF) to show how often terms show up in \textit{unique} documents. As students use various variable names, we limited the top $50$ best features (selected from a range of ($30$, $50$, $100$, $300$) in hyperparameter tuning) in TFIDF to remove redundant features. For the DKT-Expert model, two authors examined the problems in the dataset, and determined $9$ rule-based code features. These features include code component existence checks such as the usage of \texttt{else if} statements, the usage of \texttt{\&\&} operations, etc. These statements and operations represent students' usage of certain concepts such as writing alternative conditions, or using ``and" logic to solve a problem. 

% YANG - IMPORTANT - in the next paragraph, it is not clear which models you added these features to.
% Added information here, nice catch!

To improve the TFIDF and Expert models to serve as more robust baseline models, we added one additional set of features (only to baseline models) to encode information about the skills practiced in each problem, as has been done in prior work \cite{zhang2017incorporating}. Two authors examined the problem descriptions and solutions and agreed on $9$ skills we expected students to learn. For example, one skill was solving problems with negative conditions in the instructions (using words such as ``unless", ``otherwise"), requiring students to negate these conditions in their code. We represented each problem as a binary vector of practiced skills, and we used this skill vector to represent problems, instead of the one-hot encoded problem ID (see Section~\ref{sec:dkt} for model input encoding). Testing on the validation dataset showed slightly improved performance using these skill vectors.

%This design gives the baseline models access to common skills shared among problems, allowing the model to make predictions on a future problem according to skills practiced in previous problems.

\textbf{Metric}: Our primary performance metric is AUC, a standard evaluation metric for KT models \cite{piech2015deep, shin2021saint+, pandey2019self}, as it uses the predicted \textit{probability} of success, rather than a binary correctness prediction, and is more appropriate than accuracy for imbalanced datasets like ours (23\% positive).

\section{Results} % 3 pages

% \begin{table}[]
% \centering
% \caption{\label{table:first_comparison} Performance of Code-DKT compared with baseline models, predicting the first attempts of an unattempted problem.}
% \begin{tabular}{|c|c|c|}
% \hline
% Models & \begin{tabular}[c]{@{}c@{}}First Attempts\\ AUC\end{tabular} & \begin{tabular}[c]{@{}c@{}}First Attempts\\ Accuracy\end{tabular} \\ \hline
% Code-DKT                         & \textbf{75.74\%}                                             & 71.28\%                                                           \\ \hline
% DKT-TFIDF                    & 72.77\%                                                      & 67.84\%                                                           \\ \hline
% DKT-Expert                   & 69.53\%                                                      & 65.84\%                                                           \\ \hline
% DKT                          & 72.26\%                                                      & 67.67\%                                                           \\ \hline
% BKT                          & 50.22\%                                                      & 65.67\%                                                           \\ \hline
% \end{tabular}
% \end{table}

% Performance analysis
\subsection{Performance Comparison}
\subsubsection{Code-DKT vs DKT}

Table~\ref{table:assignment} shows a comparison of DKT and Code-DKT across all 5 assignments (the average of the 10 test runs). Note that for each assignment, a new model is trained and tested separately, without using data from prior assignments. This was because assignments were spaced out with weeks between then, including additional learning content, so students' performance on prior assignments is less relevant.
% This comparison shows that Code-DKT uses code features to improve the prediction performance of DKT.
To address RQ1, we consider the overall performance of the baseline DKT model on our dataset, which has an AUC of 69-75\% across assignments. This low score means it may be difficult to use model predictions to inform instruction or an automated intervention, as we discuss in Section~\ref{sec:discussion}.
To address RQ2, we see that Code-DKT \textit{consistently} outperforms DKT by 3-4\% AUC on each assignment. This shows that our approach, which augments correctness features with additional information from student code, can improve DKT predictions. For perspective, this improvement is comparable to SAINT+'s improvement over DKT on EdNet ($+2.76\%$) \cite{choi2020ednet}, or SAKT's improvement on various datasets (+3.8\%) .  

% As shown in the table, we reach two major conclusions to answer RQ1 and RQ2. First, simple code features do not improve the predictive power of DKT models. Second, with carefully designed mechanisms (such as Code-DKT), improvement is possible.

\subsubsection{Code-DKT vs Naive Code Features and BKT}
We now investigate a single assignment, A1, to illustrate Code-DKT's performance, and create a DKT-Expert baseline using assignment-specific, expert-authored code features. We selected assignment A1, as it came first (and was therefore not influenced by prior assignments) and its skills are the least complex. %because its focus on simple conditional logic makes it easy to explain case studies (Section~\ref{sec:case_study}). 
Table~\ref{table:overall_comparison} shows the performance of Code-DKT, DKT, as well as 3 new baselines: BKT, and 2 simple code-feature extensions of DKT: DKT-TFIDF and DKT-Expert (described in Section~\ref{sec:baselines}). Model performance is given for predicting all attempts (Overall) and for predicting only first attempts at each problem.
%
%The overall prediction performance results of Code-DKT compared with other baseline models are shown in Table~\ref{table:overall_comparison} (column 2). Each metric is calculated as the average of rerunning the respective models 10 times, trained on the same train-test splits.
%
The results show that neither the simple expert features nor the TFIDF data-driven features improve the overall performance of DKT. These simple features derived from student code instead negatively affect overall performance. This suggests that a more effective model structure is necessary for making use of code features, such as our Code-DKT model.
%On the other hand, Code-DKT performs $3.07\%$ better than the DKT model, showing that our code feature extraction method is able to extract information from student code and help DKT to predict student next attempt success. 
We also see that BKT has an AUC score of only 63\%, suggesting that deep models are more effective for our dataset.
%BKT has a better accuracy, but this is less meaningful, as the dataset is imbalances (with a $25.68\%$ positive rate), and it is trivial to achieve a \approx 75\% accuracy.

 \subsubsection{When is Code-DKT Effective?}

We used assignment A1 to investigate \textit{when} Code-DKT was more effective than DKT, helping to answer RQ3.
 
\textbf{Overall vs First Attempts}: We investigated Code-DKT's  performance at predicting a student's \textit{first} attempt at each problem (Table~\ref{table:overall_comparison}, column 3).  First attempts are important in a KT task because they represent points at which an ITS might make key interventions (e.g. offering a worked example if a student might fail at problem solving). Therefore, many KT evaluations differentiate a student's first attempt on a task (where a model must make predictions using only performance on \textit{other} problems) from subsequent attempts. This distinction also helps us understand when the Code-DKT model is most effective. One might ask, is Code-DKT using student code submissions to learn a better representation of student knowledge (which \textit{would} help it predict first attempts), or is it simply estimating how close a student is to solving the \textit{current} problem (which would only help to predict \textit{subsequent} attempts). Our results shows that Code-DKT actually performs \textit{best} when predicting first attempts, and it also shows a similar improvement over DKT for first attempts (+$3.48\%$), compared to all attempts (+2.93\%) . This suggests that the content of a student's code is helpful for not only predicting how quickly they will solve the \textit{current} problem, but also \textit{future} problems.
%Simple code features in this situation have similar (DKT-TFIDF) or lower (DKT-Expert) performance than DKT, consistent with the observation of overall predictions, suggesting that such feature are similarly unhelpful for predicting first attempts.

\begin{figure*}
\centering

\begin{minipage}[b]{0.47\textwidth}
\includegraphics[width=\textwidth]{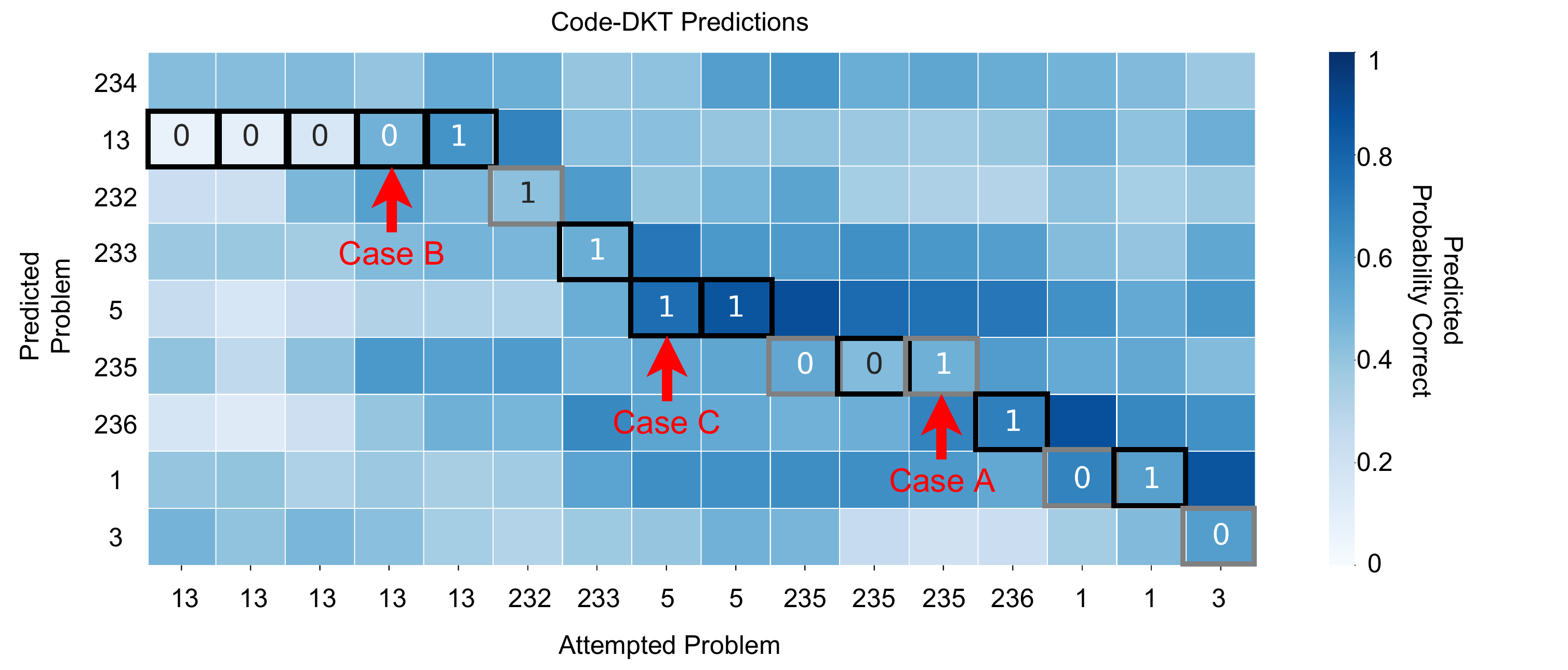}
\caption{\label{fig:Code-DKTheatmap} Code-DKT generated correctness predictions heatmap for a student.}

\end{minipage}
\hfill
\begin{minipage}[b]{0.47\textwidth}
\includegraphics[width=\textwidth]{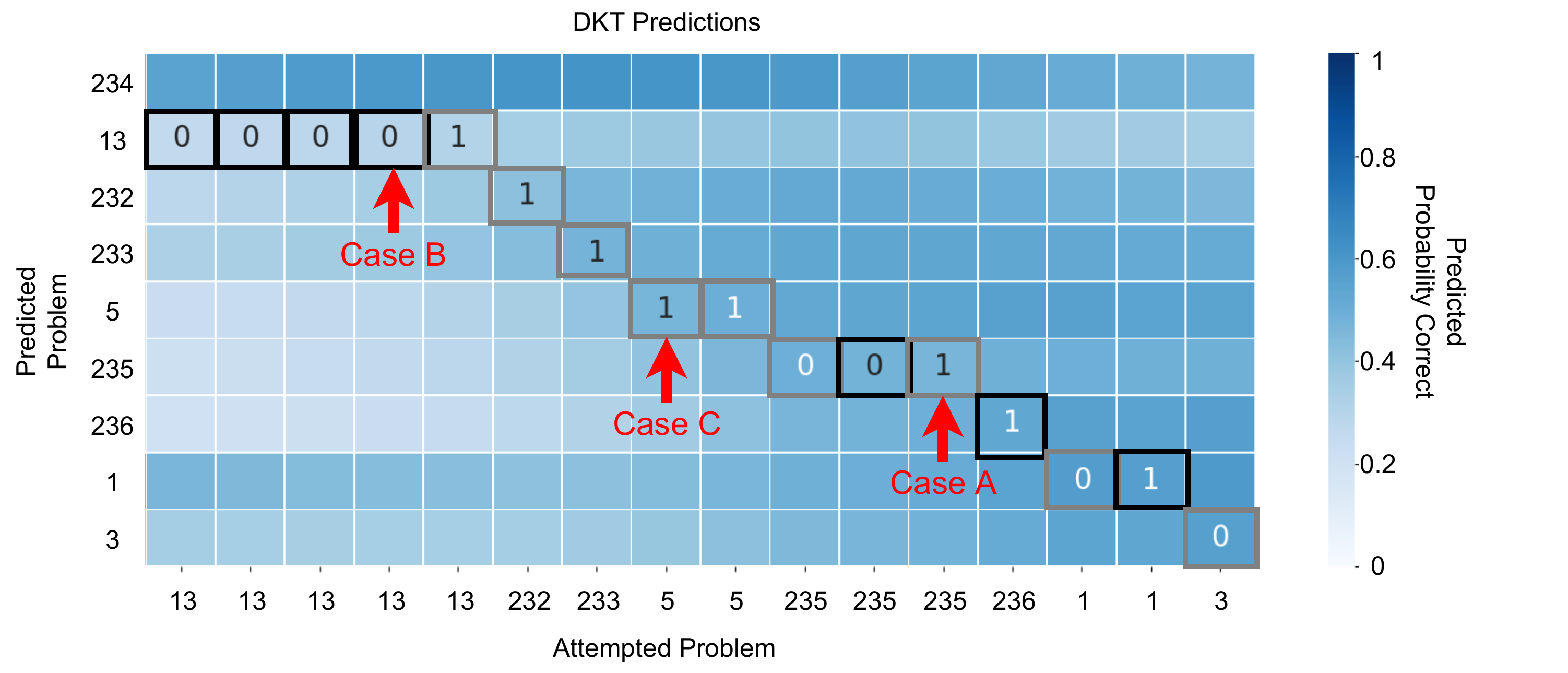}
\caption{\label{fig:dktheatmap} DKT generated correctness predictions heatmap for a student.}

\end{minipage}
% \vspace{-0.2cm}
\end{figure*}

% \begin{figure*}
% \centering

% \begin{minipage}[b]{0.49\textwidth}
% \includegraphics[width=\textwidth]{Figures/DCKTHeatmap2.pdf}
% \caption{\label{fig:Code-DKTheatmap2} Code-DKT generated correctness predictions heatmap for Student B.}
% \end{minipage}
% \hfill
% \begin{minipage}[b]{0.49\textwidth}
% \includegraphics[width=\textwidth]{Figures/DKTHeatmap2.pdf}
% \caption{\label{fig:dktheatmap2} DKT generated correctness predictions heatmap for Student B.}
% \end{minipage}
% % \vspace{-0.2cm}
% \end{figure*}

% \subsubsection{Problem-Specific Results}

\textbf{Problem-specific Performance}: Table~\ref{table:decomposition} shows the decomposed AUC performance of Code-DKT and DKT on each problem. We observe that Code-DKT outperforms DKT overall on $6$ of the $9$ problems. The difference ranges from +15.54\% AUC (problem 13) to -4.43\% (problem 236), suggesting that the benefit of Code-DKT's code features depends somewhat on the programming problem. It also shows that code features \textit{can} reduce model performance, but the potential for Code-DKT's improvement seems to be greater than the potential for harm.

%On some problems, for example problem $13$, the overall prediction AUC of Code-DKT is $11.96\%$ higher than DKT, while when predicting a new problems, Code-DKT over-performs with a margin of $15.54\%$. However on the problem $236$, Code-DKT falls behind by $4.43\%$ on the overall performance. This result also shows that the Code-DKT method would perform better on certain problems, but not necessarily on an another problem.

% A closer look on the problem descriptions on two cases of students will be given in Section~\ref{sec:case_study}. Here we give the problem instructions of Problems $13$ and $236$.\footnote{We will release problem instructions for these $9$ problems after the review period for reference.}

% \textbf{Problems $13$}: Write a function in Java that implements the following logic: You are driving a little too fast, and a police officer stops you. Write code to compute the result, encoded as an int value: 0=no ticket, 1=small ticket, or 2=big ticket. If speed is 60 or less, the result is 0. If speed is between 61 and 80 inclusive, the result is 1. If speed is 81 or more, the result is 2. Unless it is your birthday--on that day, your speed can be 5 higher in all cases.

% \textbf{Problems $236$}: You have a green lottery ticket, with ints a, b, and c on it. If the numbers are all different from each other, the result is 0. If all of the numbers are the same, the result is 20. If two of the numbers are the same, the result is 10.

To understand \textit{when} Code-DKT's code features were useful, we investigated differences between the problems where it outperformed DKT and those where it did not. We found that many of the problems where there was improvement shared similar learning concepts and solution structure. For example, problems 3, 232 and 234 all used the ``independent choice'' programming pattern, which is often solved with nested if-statements. Similarly, problems 1, 3, 5 and 13 all included a pattern where one condition changes a value used in another condition. These common patterns seem to have helped the model make better predictions on problems that used them. However, 2 of the 3 of the problems where Code-DKT performed poorly involved a unique learning concept that did not appear in any other problems. For example, problem 236 requires students to check if any 2 of the 3 given variables are equal (which has no analog among other problems) and 233 requires the \texttt{Math.abs} function (which many students failed to use correctly). Together, these results suggest a hypothesis that Code-DKT's code features are most useful at predicting problems that share code structures with other problems, and less useful at predicting problems that emphasize novel code structures. This suggests Code-DKT may be successfully modeling students' knowledge of common code patterns.%, which can be generalized across problems.

\begin{table}[]
\centering
\caption{\label{table:decomposition} Decomposed performance of Code-DKT and DKT AUC performance on different problems in assignment A1.}
\begin{tabular}{|c|cc|cc|}
\hline
\multirow{2}{*}{Problems} & \multicolumn{2}{c|}{Code-DKT}                            & \multicolumn{2}{c|}{DKT}                                 \\ \cline{2-5} 
                          & \multicolumn{1}{c|}{Overall}          & First            & \multicolumn{1}{c|}{Overall}          & First            \\ \hline
234                       & \multicolumn{1}{c|}{\textbf{64.60\%}} & 71.38\%          & \multicolumn{1}{c|}{63.75\%}          & \textbf{73.48\%} \\ \hline
13                        & \multicolumn{1}{c|}{\textbf{78.45\%}} & \textbf{86.55\%} & \multicolumn{1}{c|}{63.59\%}          & 68.81\%          \\ \hline
232                       & \multicolumn{1}{c|}{\textbf{74.93\%}} & \textbf{78.99\%} & \multicolumn{1}{c|}{72.49\%}          & 73.09\%          \\ \hline
233                       & \multicolumn{1}{c|}{64.79\%}          & 74.57\%          & \multicolumn{1}{c|}{\textbf{67.18\%}} & \textbf{76.33\%} \\ \hline
5                         & \multicolumn{1}{c|}{\textbf{75.38\%}} & 81.34\%          & \multicolumn{1}{c|}{74.28\%}          & \textbf{81.79\%} \\ \hline
235                       & \multicolumn{1}{c|}{70.65\%}          & \textbf{71.96\%} & \multicolumn{1}{c|}{\textbf{75.03\%}} & 70.80\%          \\ \hline
236                       & \multicolumn{1}{c|}{74.25\%}          & 74.30\%          & \multicolumn{1}{c|}{\textbf{78.68\%}} & \textbf{77.06\%} \\ \hline
1                         & \multicolumn{1}{c|}{\textbf{68.62\%}} & 70.32\%          & \multicolumn{1}{c|}{66.67\%}          & \textbf{73.20\%} \\ \hline
3                         & \multicolumn{1}{c|}{\textbf{71.00\%}} & \textbf{71.00\%} & \multicolumn{1}{c|}{64.02\%}          & 64.02\%          \\ \hline
\end{tabular}
\end{table}

\subsubsection{Ablation Study}
\label{sec:ablation}

Our Code-DKT model design choices include: where to incorporate correctness information, how to update the embedding, and what underlying network to use (LSTM or RNN). Table~\ref{table:ablation} shows the results of an ablation study on assignment A1 to determine which of these choices improved the performance of our final DKT model (first row). 
The final Code-DKT model concatenates the \textit{correctness} of a students' attempt with code features in two places (see Section~\ref{sec:code-dkt}): before the attention mechanism (the vector $\mathbf{e}_r$), and in the final trace fed into the LSTM ($\mathbf{z_i}$ concatenated with $\mathbf{x_i}$). The model in row 2 only includes correctness information in the first case, and row 3 includes it only in the second case. Both models lose performance, but not by much ($0.5\%$), suggesting that correctness information helps both in attending to relevant code paths, and final predictions, but this information is somewhat redundant.
We also investigated using an RNN (row 4) instead of an LSTM, but this was, as predicted, moderately less effective.
Finally, recall that Code-DKT uses code2vec to embed students' code as a vector, and updates this embedding throughout model training. Row 5 shows a version where we pretrained this embedding on the training dataset, using code2vec to predict the correctness of students' code, and then fixed the embedding when training the LSTM. This model does much worse, suggesting that the relevant features for predicting the \textit{correctness} of code are different from those for predicting \textit{future performance}.

\begin{table}[]
\centering
\caption{\label{table:ablation} Code-DKT ablation study on A1.}
\begin{tabular}{|c|l|c|c|}
\hline
%& Models              & \begin{tabular}[c]{@{}c@{}}Overall\\ AUC\end{tabular}  \\ \hline
& Model    & Overall AUC \\ \hline
1 & Code-DKT (Final Model)               & \textbf{74.31\%}                                                                             \\ \hline
2 & Correctness: Attention Only     & 73.81\%                                                                                                  \\ \hline
3 & Correctness: Trace Only & 73.84\%                                                                                                   \\ \hline
4 & Model: RNN           & 73.63\%                                                                                                  \\ \hline
5 & Embedding: Static           & 68.74\%                                                                                                  \\ \hline
\end{tabular}
\end{table}

% Closer look on different problems
\subsection{Case Studies}
\label{sec:case_study}

% Heatmap analysis
To further answer RQ3, we examined \textit{how} code features may have improved Code-DKT through $3$ case studies. We use prediction heatmaps from Code-DKT and DKT for one student, shown in Figures~\ref{fig:Code-DKTheatmap} and \ref{fig:dktheatmap} for Code-DKT and DKT, respectively. 
%while Figures~\ref{fig:Code-DKTheatmap2} and \ref{fig:dktheatmap2} are the heatmaps for student B.
The rectangular cells show which problem the student actually attempted (y-axis) at each time-step (x-axis), and the numbers in the cells represent the ground truth values of whether student's attempt was successful (1) or unsuccessful (0). Black frames indicate correct (i.e. accurate) model predictions, while grey ones indicate incorrect predictions. The color of the heatmap in each cell specifies the predicted probability of students making a correct submission on a given problem (y-axis) at the given time-step (x-axis), and darker means a higher probability of success. For example, in Figure~\ref{fig:Code-DKTheatmap}, the student makes 4 unsuccessful attempts at problem 13, followed by a successful attempt, then succeeds at problems 232 and 233 in one attempt each.
% The X-axis is the sequence of the next problem IDs from their submissions, where for example, on the heatmaps from student A, the leftmost cells are the predicted probabilities of students making their first submission correct, and the problem they submitted was Problem $13$ (they passed $234$ with one attempt).

The heatmaps for the student (Figures \ref{fig:Code-DKTheatmap} and \ref{fig:dktheatmap}) show that Code-DKT is able to make better predictions on the traces than DKT, making $11$ out of  $16$ successful predictions, while DKT is able to make $8$ of them correct. Another observation is that Code-DKT heatmaps have much stronger predictions with values close to $1$ or $0$ compared with DKT, showing that with code features, the model is more confident.

\label{sec:casea}

\begin{figure}
\begin{center}
\includegraphics[width=0.47\textwidth]{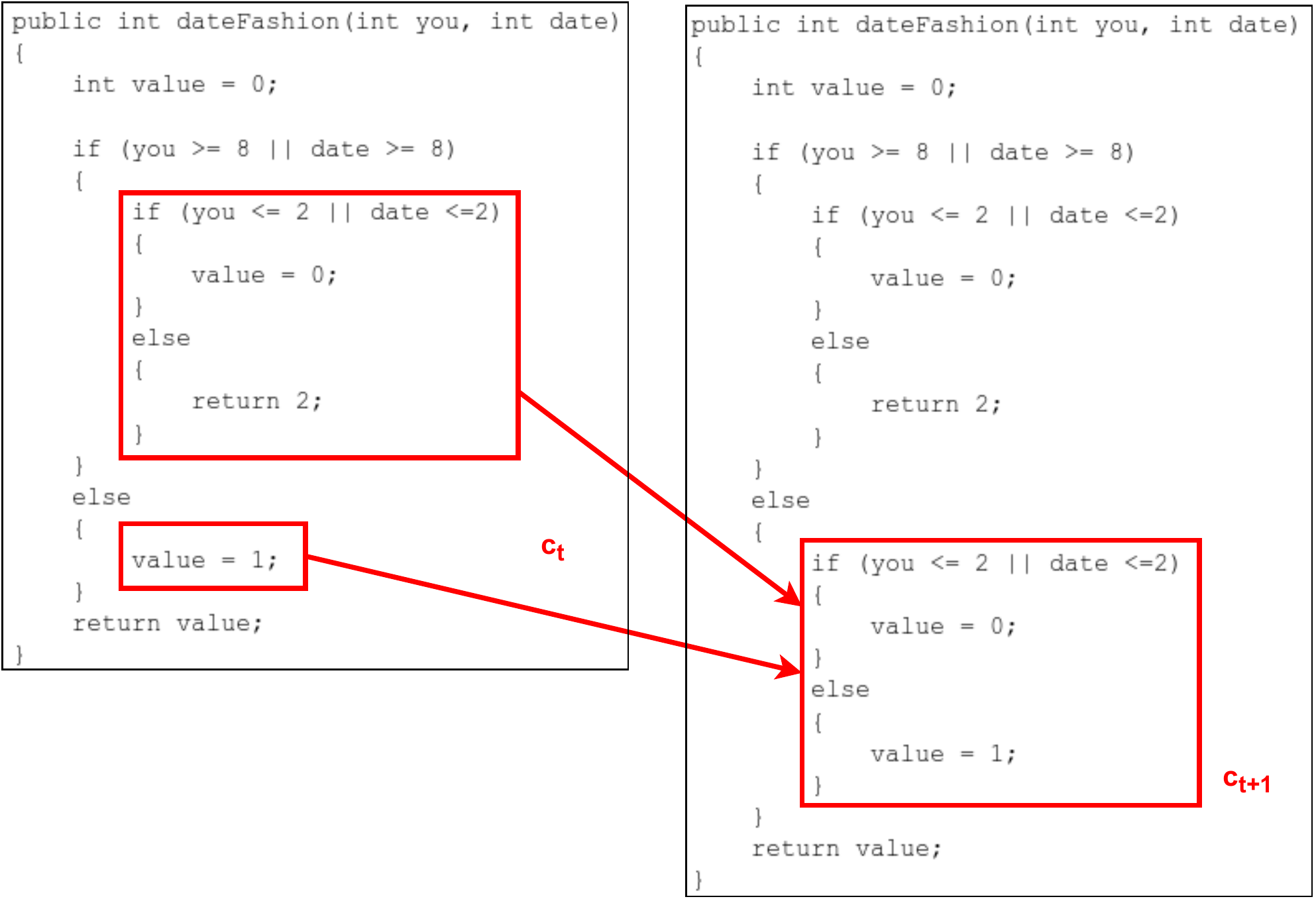}
\caption{\label{fig:casea} Code at times $t$ and $t+1$ for Case A, where the code $c_1,...c_t$ is used to predict correctness at $t+1$.}

\end{center}
\end{figure}

% When code works
\textbf{Case A: Successful Prediction}:
In Case A, Code-DKT uses code features to make better predictions than DKT on the predictions of the student's final submission on Problem $235$. As shown in Figures~\ref{fig:Code-DKTheatmap} and \ref{fig:dktheatmap}, while both Code-DKT and DKT can successfully predict the incorrect submission on the student's second submission of Problem $235$ and fail to predict the correct submission on the third, Code-DKT gives a higher prediction than DKT. In Figure~\ref{fig:casea}, the student's code submissions show the reason. The student's second submission is almost correct, demonstrating a correct (if inefficient) nested if-else structure, but they have omitted the nested condition in their else branch. Code-DKT is able to infer the quality of the student's code, since its prediction of success probability increased from 44.2\% to 49.1\% after the student's second (incorrect) attempt, while DKT's prediction decreased from 47.7\% to 46.7\%. Code-DKT's higher prediction may be because the if-else structure the student was missing was very similar to one they had already written, as shown in Figure~\ref{fig:casea}. These code structures are easily captured by the path-based AST representation used by code2vec.
%
% Problem $235$ requires a nested \texttt{if} implementation with relatively complex logic design. Both levels of \texttt{if} conditions depends on the values of \texttt{you} and \texttt{date} variables. The outer \texttt{if} checks if any of the variables are $>=8$, while the inner \texttt{if} checks if any of the variables are $<=2$. The code submission $c_t$ is incorrect, because the logic in the outer \texttt{else} is wrong. It should have checked the $<=2$ conditions of either variables again to make a correct submission, but failed to do so. Though, the student is able to make a correct submission on $t+1$ with the submission $c_{t+1}$. 
%
Without code features, it is difficult for DKT to predict whether the student is going to succeed on $t+1$, since it only knows the student has failed twice, not how close they are to succeeding. Even with code features, there is still a great deal of uncertainty. No matter how close a student is to a correct answer, there is no guarantee they will achieve it on their next attempt. This may help to explain why Code-DKT does not more dramatically outperform DKT overall.

%$c_t$ has all the pieces ready to build a correct submission. First, it has the correct inner \texttt{if} logic, specified in the upper red frame of $c_t$. The student should again check this condition in the \texttt{else} of the outer \texttt{if}. Second, they would need a change of the return value of the inner \texttt{else} condition. This only change they need is already ready for them at time $t$, specified in the lower red frame. In the \texttt{else} condition of inner \texttt{if}, the return value should be $1$ instead of $2$. The student probably knows it, but just thought the value could be directly assigned when outer \texttt{if} condition is false. Code-DKT catches the code features from both pieces, and predicts a possible success on their next attempt. Although Code-DKT is able to make a correct prediction on $t+1$ for Problem $235$, it could as well be hard to know how long it will take before Student A realizes the logic issue. However, given a trace of submission history, the student does not typically take a lot of submissions before they succeed for similar problems (e.g. Problems $232$, $5$, see Section~\ref{sec:caseb} about Code-DKT making correct predictions on them), Code-DKT gives a positive prediction, though not very strong. 

\label{sec:casec}
% Originally Case C!
\begin{figure}
\begin{center}
\includegraphics[width=0.40\textwidth]{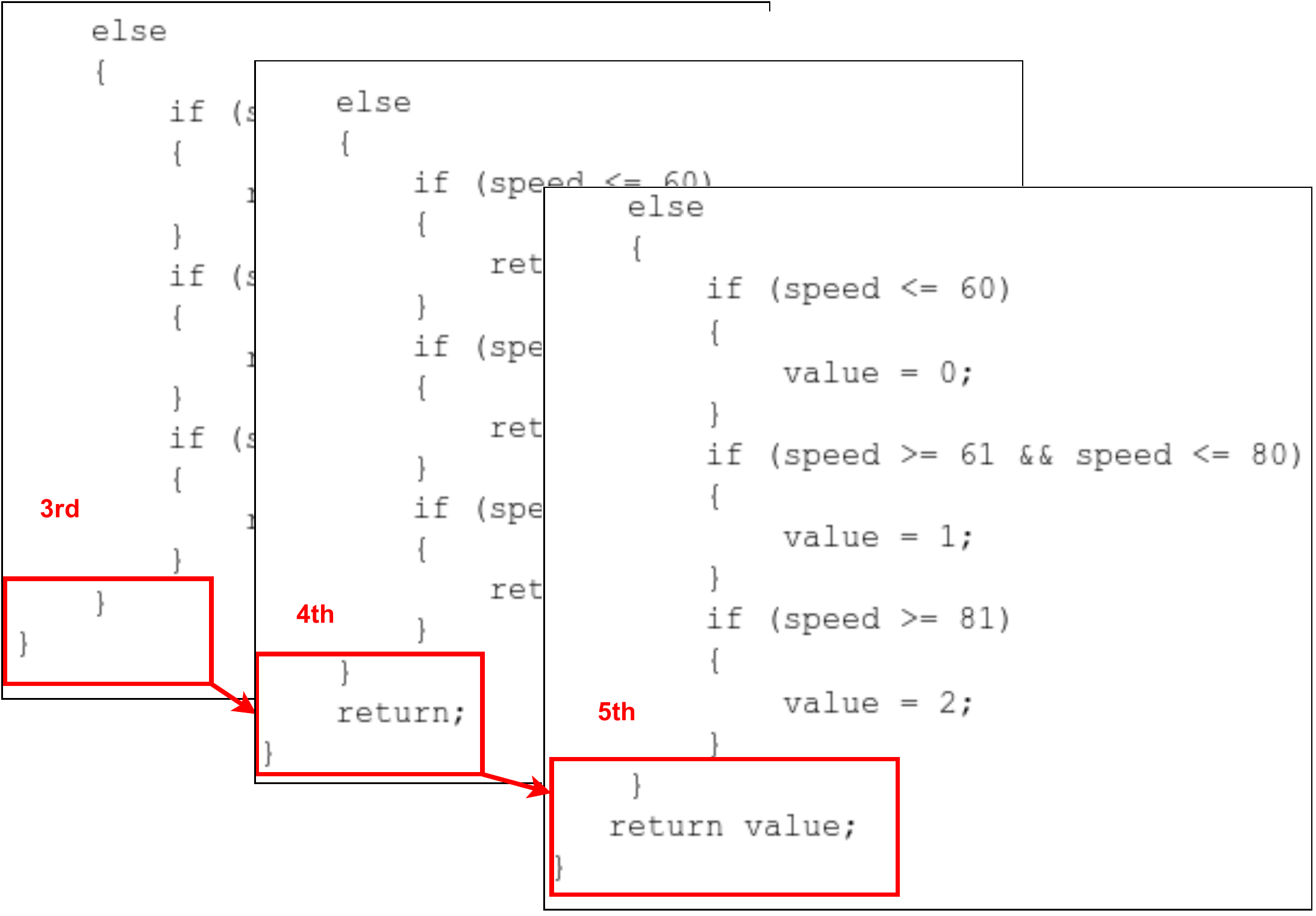}
\caption{\label{fig:casec} Case B 4th, 5th and 6th code attempts. }

\end{center}
\end{figure}
\textbf{Case B: Unsuccessful Prediction}:
% When code makes prediction wrong
Case B shows that even when a student's code is nearly correct for a given problem, it doesn't guarantee that they will be successful on their next attempt. Sometimes Code-DKT is overconfident in these situations, and incorrectly predicts success, as in Case B. Figure~\ref{fig:casec} shows the last three attempts the student made on Problem $13$: two incorrect followed by a final correct attempt. The only differences between the final attempt and the earlier two is shown in the red frames. The student's 4th attempt achieved the correct logic for Problem 13, the 5th attempt adds an empty \texttt{return} statement, and the 6th and final attempt adds the appropriate return value. After seeing the almost-correct code at their 4th attempt, Code-DKT predicted that the student would succeed on the next attempt since the modifications they needed were minimal (just write ``\texttt{return value;}"), but it took one extra attempt to get it right. An expert might make a similar conclusion, that the student was close enough to realize their mistake and submit a correct answer, and would have similarly been wrong. This highlights the uncertainty present in any KT task and the challenges of applying KT to student code.

\label{sec:caseb}

\begin{figure}
\begin{center}
\includegraphics[width=0.47\textwidth]{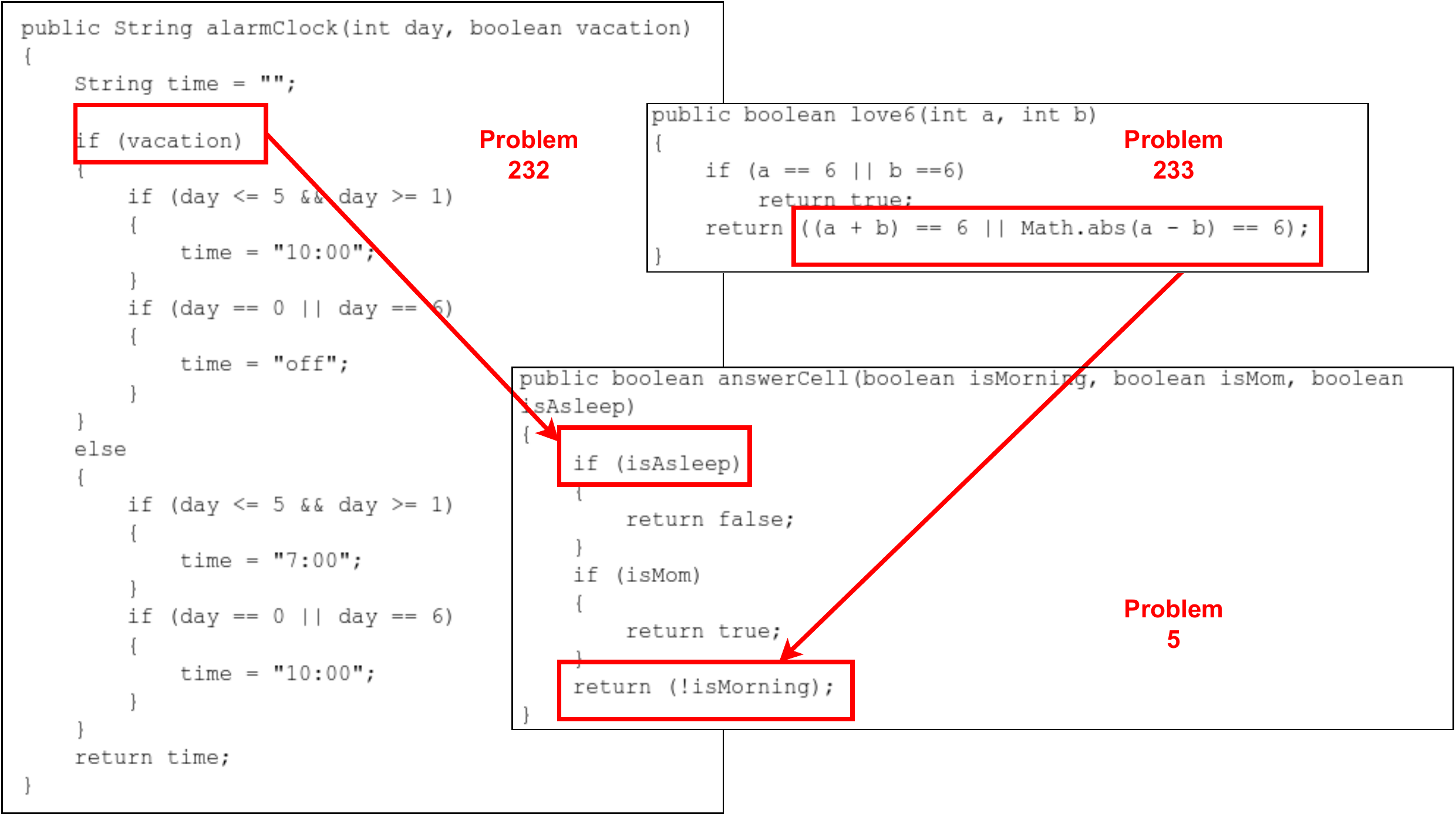}
\caption{\label{fig:caseb} Case C, using code from problems 232 and 233 to predict the same student's performance on problem 5}

\end{center}
\end{figure}
\textbf{Case C: Successful Prediction (First Attempt)}:
Case C illustrates that Code-DKT can also use code from \textit{previous problems} to improve its predictions of \textit{first attempts} of new problems (as shown quantitatively in Table~\ref{table:overall_comparison}). For example, when the student successfully completes problems 232 and 233 in a single attempt, Code-DKT's prediction of the student's success on problem 5 increases from 42.0\% to 50.0\% to 72.5\% respectively, leading it successfully predict success on the student's first attempt at problem 5. However, DKT's confidence only modestly increased from 41.8\% to 47.3\% to 47.0\%, leading it to incorrectly predict failure. Both models know that these problems are related, and share some learning concepts (based on how other students' successes on the problems are related), but Code-DKT's analysis of the student code allowed it to infer more about the knowledge that was demonstrated in past problems.

% When first attempt works 
% For Student A, the first attempt of Problem $5$ is successfully predicted, as the student has successful submissions on $13$, $232$, $233$. While Code-DKT is able to make correct and confident predictions that the student will succeed on their first tries of $233$ and $5$, the DKT model gives wrong predictions. This shows that Code-DKT is able to make more accurate predictions on first attempts with code features.

We use these three consecutive code submissions to explain why this may be the case in Figure~\ref{fig:caseb}. 
%The attempts $c_{t-1}$ and $c_{t}$ are the only and correct submissions of Problems $232$ and $233$ respectively, used to predicting the first attempt on Problem $5$. We also show $c_{t-1}$ since Problem $233$ is a different type than $232$ and $5$, as any of four conditions for a single \texttt{if} condition would give an return value of $1$. 
%
%Compared to DKT, Code-DKT gives very strong predictions that the first attempt of Problem $5$ would be a successful one. Despite the problem itself is simple, some evidences can also be seen from $c_{t-1}$ and $c_{t}$. 
For example, in Problem $232$, the student directly uses Boolean variable in the if-condition (\texttt{if (vacation)}) rather than a superfluous comparison (\texttt{if (vacation == true)}) that many students use, demonstrating a higher level of understanding. This same direct usage of Boolean variables is seen in the \texttt{if} condition and \texttt{return} statement of Problem $5$. The code submission on Problem $233$ further suggests the student is able to combine logical operators with Boolean variables to \texttt{return} a Boolean expression. This occurs again in the \texttt{return} statement of the students' attempt at Problem $5$, shown in the lower rectangular. While we cannot know for certain which code features Code-DKT used to make its success prediction for Problem $5$, these repeated code structures are one possibility, given code2vec's ability to recognize repeated patterns in ASTs.

\section{Discussion} % 1 page
\label{sec:discussion}

\textbf{RQ1: How well do domain-general models perform?} We used domain-general KT models (DKT) as the baseline models for our programming dataset. These models performed relatively poorly, averaging 73.09\% AUC across assignments. While this is considerably better than chance, the performance may not be high enough to use in some student modeling contexts. For example, for assignment A1, the recall of DKT was 31.4\% and the precision was 46.5\%, so the model fails to identify two thirds of unsuccessful attempts, and over half of the time when the model predicts a failed attempt, the student actually succeeded. This suggests that KT is a difficult challenge on this dataset. By contrast, DKT has historically been effective on other datasets, which are both larger and in other domains, such as EdNet \cite{choi2020ednet}, Assistments \cite{selent2016assistments} and KhanAcademy \cite{piech2015deep}. One possibility is that the more complex nature of programming problems, with myriad possible correct and incorrect solutions, makes KT prediction more challenging on this dataset, compared to those in other domains. If this is the case, several aspects of programming may contribute to the challenge of modeling student success. Programming problems often require many attempts to get correct (6.1 on average in our dataset), leading to class imbalance. In our dataset, the problem descriptions were complex, and their solutions involved complex conditional logic,  and students had to write perfect Java syntax for the program to compile. These factors mean there are many ways for students to make small ``slips", making the relationship between skill and success less direct. 
%In addition, we track the binary correctness, which does not include the detailed scores of how good students have done in their attempts as well. These factors together made this KT task specifically hard, and reflected in the baseline models.

Another possibility is that our dataset (410 students) was simply too small for complex deep models to find success, compared to the 1000s or even 100,000s of learners in other datasets where DKT has been evaluated. However, model complexity alone does not explain the difference, since the simpler BKT model did even worse than DKT, and our Code-DKT model, which had far more parameters, performed better. Additionally, DKT has historically performed well on some other small datasets (e.g. the ``ASSIST-Chall'' and ``STATICS'' datasets from \cite{pandey2019self} with 300-700 students). Regardless, many tutoring systems only have hundreds of students, and effective KT models must still be able to perform well on these small datasets. Thus, to the extent that our datasets is representative of the domain, our results suggest the need for improved KT models for programming.

\textbf{RQ2: How can code features improve KT models?} Our results show that a simple extension of DKT with code features does not improve its performance. This result is somewhat surprising, given that relatively simple features (e.g. the presence of a \texttt{return} statement) should be at least somewhat related to how close a student is to a correct answer. It is possible such features may improve a model with different structure, but in our dataset, they were not helpful to DKT. This suggests the need for thoughtful approaches to incorporating domain-specific features into deep models. Our Code-DKT model was able to make reasonable improvements to DKT (+3.07\% overall on A1). This is comparable to the improvement of SAINT over DKT on the EdNet dataset (they achieved +2.76\% in AUC), or SAKT over DKT on various datasets (+3.8\%) \cite{pandey2019self}. This suggests that domain-specific features can be just as important as model structure for effective KT. Code-DKT's improvement is also robust. It has a +3\% to +4\% improvement overall on all five assignments. Importantly, however this is still a relatively poor performance overall, suggesting the need for more work on leveraging domain-specific features for improved KT.

\textbf{RQ3: When and how do code features work?} We also explored when and how the code features improved model performance. We found that code features are most useful on problems that share similar learning concepts with other problems in the dataset, and less useful on problems with unique and difficult concepts (e.g. \texttt{Math.abs()}). This makes sense -- if we make an analogy to the original BKT where each problem was labeled with KCs, if you had a unique KC, the model would have no way of predicting on that problem. In our case, the KCs are inferred by the model, but the same limitation exists. However, most problems in our dataset did share primary learning concepts (e.g. loops, conditionals) and benefit from code features, and this repeated practice is a common feature of many CS1 courses. We also found that code features are useful for predicting both first attempts and subsequent attempts. Our case studies reveal potential mechanisms for both of these effects. For repeated attempts, the model seems to use the relative correctness of a student's code to determine how close they are to a solution and therefore how likely they are to get it right on the next attempt. For first attempts, the model seems to identify code structures in prior attempts that indicate knowledge or competence with certain programming concepts, which it uses to make predictions on new problems. More work is needed to verify these hypotheses, and to understand how the model represents this knowledge.

\textbf{Limitations}: Our model and experiment have several limitations. 1) All models evaluated, including Code-DKT, have a relatively low performance, partially due to the difficulty of the problem and low data size ($410$ students), as discussed in Section~\ref{sec:discussion}. Still, they perform considerably better than chance, and such models could still be useful, e.g. in prioritizing help to struggling students.
%However, while the benchmark for EdNet is currently 79.14\% AUC with more than $780$k students on multiple choice problems, our model could improve when more data is available. 
% 2) While we only have an improvement of $3.07\%$ on AUC, it is comparable to SAINT+ \cite{shin2021saint+} improvement to DKT on EdNet (+2.76\%) already. 
2) Our dataset was from a single semester of a course. While our semester-long dataset of 50 problems is considerably more robust than some of the prior work on KT in programming (e.g. using 1-2 problems \cite{wang2017learning}), it is unclear how our results will generalize to  other semesters, classes or programming languages.
%3) We evaluated our model on problems that focused on a single concept (conditionals), and our model may be most applicable to problems sharing similar structures, as it relies on the code features. We have not yet applied it to a course-level dataset to verify the performance when problems cover more concepts.
3) We used only DKT as a baseline model to extend and to compare against, and it is possible code features may have different effects on other models. However, as explained in Section~\ref{sec:dkt}, DKT has a comparable performance to more modern deep models, and made sense as a starting point to explore the effect of code features. 
% 4) We used code2vec as the base model of code features extraction, just to show that domain-specific code features extracted with structural information helps the performance of KT for programming education. Further investigations could be given to other models for code feature extraction as well.
% Low performance, but EDnet too!

% Might be more applicable to consistent concepts or structures

% Did not try other code feature extraction algorithms, as we just want to show code features can help it!

% Small data size with 363 students

\section{Conclusion} % 0.25 page
The contributions of the paper are 1) the Code-DKT model, which extends DKT with embedded code feature extraction; 2) results showing that CodeDKT consistently improves over DKT in a programming dataset; and 3) comparisons and case studies highlighting when and why Code-DKT code features help. This paper compared our new Code-DKT model to domain-general BKT and DKT baselines, and two DKT models extended with simple code features, demonstrating improved performance for Code-DKT over these baselines. However, the best baseline model performance was about 73\%, and Code-DKT was 74.3\%, demonstrating considerable room for improvement on modeling for knowledge tracing in programming. The case studies in this paper illustrate specific situations where knowledge tracing can be particularly difficult in programming, and where there is potential for improving code KT, e.g. when common code structures are used across problems.

%ACKNOWLEDGMENTS are optional
\noindent \textbf{Acknowledgements:} This material is based upon work supported by NSF under Grant No. \#2013502.

%
% The following two commands are all you need in the
% initial runs of your .tex file to
% produce the bibliography for the citations in your paper.
\bibliographystyle{abbrv}
\bibliography{yang}  % sigproc.bib is the name of the Bibliography in this case

\end{document}